\documentclass[11pt]{article}
\usepackage{moriond,epsfig}

\def\be{\begin{equation}}
\def\ee{\end{equation}}
\def\bea{\begin{eqnarray}}
\def\eea{\end{eqnarray}}

\def\gsim{~\rlap{$>$}{\lower 1.0ex\hbox{$\sim$}}}
\begin{document}
\vspace*{4cm}
\title{Relating leptogenesis and dark matter}

\author{ Michel H.G. Tytgat }

\address{Service de physique th\'eorique CP225\\
Universit\'e Libre de Bruxelles, Belgium}

\maketitle\abstracts{
A scenario that  relates the abundance of dark matter to the baryon asymmetry of the Universe is presented.
In this scenario, based on a left-right extension of the Standard Model, dark matter is made of light, $M\sim 1$ GeV, for all
practical matters non-interacting, right-handed Majorana neutrinos.}

\section{Introduction}

Only one fifth of matter in the universe is made of atoms, according to the so-called "concordance model" of cosmology.
These atoms, or baryons, are thought to have survived to annihilation with the anti-matter that filled the early universe
thanks to the existence of a primordial baryon asymmetry.
This baryon asymmetry, in turn, could have been generated dynamically, typically in the very early universe, following a scheme known as baryogenesis.

The rest of matter in the universe is supposedly made of Dark Matter, a speculative component proposed already many decades ago
in order to explain the dispersion of
velocities of galaxies bound in clusters.
The simplest model of dark matter posits the existence of a thermal relic of stable, weakly interacting massive particles or WIMPs. Those were in thermal
 equilibrium in the early universe, but then fell out of equilibrium, leaving an abundance that is simply related to their annihilation cross-section.
Dark matter of that kind is already expected to exist. We believe that there is a cosmic background of Standard Model (SM) neutrinos, that
 decoupled around $T \sim 1$ MeV when the weak interactions fell out of equilibrium, incidentally a first step toward the synthesis of the lightest elements.
Those neutrinos are massive, but there are not heavy enough to be the dominant form of dark matter in the universe.
Large-scale structures formation lore requires dark matter to consist of (or to be equivalent to) particles that are heavier than $\sim 100$ eV, while the
neutrinos we know of have a mass of at most $\sim 1$ eV. Hence there must be something else.
The most popular, well-motivated, and indeed quite successful candidate for dark matter beyond the SM is a neutralino, the lightest, preferably neutral,
 of the supersymmetric partners of the known particles.

 This might be all correct. Yet, there is a generally overlooked
 puzzle posed by these somewhat standard approaches to the matter problems of the universe.
 If baryons ought their survival to baryogenesis and dark matter its existence to freeze-out, phenomena that, supposedly, took
 place at very different times in the early universe, how come that their abundance are so similar today? Notice that while the ratio
 \begin{equation}
\label{ratio}
 {\Omega_{dm}/ \Omega_{b}} \sim {5}
\end{equation}
is preserved by the expansion of the universe nowadays, this what not the case after freeze-out but while the baryons were still relativistic
 \begin{equation}
\label{ratiobefore}
 {\Omega_{dm}/ \Omega_{b}} \propto {a}
\end{equation}
where $a$ is the scale factor. Albeit not the most dramatic, this is one of the many adjustments or fine-tuning problems posed by our (understanding of
our) universe.

It is possible that the explanation for (\ref{ratio}) is  anthropic in origin \cite{Tegmark:2005dy}. It could also be mere serendipity.
For instance, leptogenesis, a possible
mechanism for the origin of the baryon asymmetry, requires neutrinos to be massive, a requisite for dark matter. Things
could hardly be simpler (a missed opportunity for our universe)
 but, yet again, the known neutrinos can not be the dominant form of dark matter. In the present proceedings, we will
report on a recent attempt to relate the baryonic and dark components using (a)symmetry principles \cite{Cosme:2005sb}. Specifically, we shall relate the baryon asymmetry to a corresponding
asymmetry
in the dark matter. This idea is not quite knew \cite{Barr:1990ca,Kaplan:1991ah,Kuzmin:1996he} but, as usual, the devil is in the details.
We develop  a scenario, based on the more recent work of Kitano and Low \cite{Kitano:2004sv},
which, for lack of something better, one could  dub "Matter Genesis", since both baryonic and dark matters have to be generated.

\bigskip

\section{Outline of the model}

We distinguish a visible and a dark sector. The SM particles belong to the visible sector, but we will also add other
states. As usual, the lightest particle of the dark sector is
protected from decay by a discrete symmetry, analogous to R-parity.
In our scenario, this dark matter candidate is a right-handed
neutrino that we call the $\nu_R$. 

In our model, based on the gauge group $SU(2)_L \times SU(2)_R \times U(1)_{B-L}$, the $\nu_R$ interact through the exchange of 
heavy  $SU(2)_R$ gauge bosons. Now we would like the abundance of $\nu_R$ to be similar to that of baryons $n_{\nu_R} \sim n_b$. This constraint implies that the the mass
of the $\nu_R$ is $\cal O$(GeV) to account for the dark matter energy density, a mass scale much smaller than that required by thermal leptogenesis. 
Second, as the $\nu_R$ are very weakly interacting, the abundance of $\nu_R$ can not be thermal 
since otherwise $n_{\nu_R} \sim n_\gamma$.  To circumvent 
these tensions, we introduce, following Kitano and Low, a odd-parity "messenger particle", that we call the $D$.
 In our model, the $D$ are colored, $SU(2)$ singlets states, and they carry a $B-L$ number. A net $D$ particle $B-L$ asymmetry will be produced at
 some high scale, together with a corresponding $B-L$ asymmetry in the even-parity sector, through a standard thermal leptogenesis mechanism for which 
 we need extra heavy Majorana states.
 
 A model which contains all these ingredients has been proposed  by Davidson {\it et al} 
\cite{Davidson:1987mh} as an alternative way of giving
mass to the quarks and leptons. It is known in the literature as
the "universal see-saw model". The gauge group is $SU(2)_L\times
SU(2)_R\times U(1)_{B-L}$. The left and right-handed quarks
$Q_{R,L}$ and leptons $L_{R,L}$ are respectively $SU(2)_L$ and
$SU(2)_R$ doublets and, in the simplest framework, there are two
Brout-Englert-Higgs (BEH) doublets,
$$\phi_L \sim (2,1,1)$$
 and
 $$\phi_R \sim (1,2,1).$$
  To give mass to the
quarks and leptons, one introduces a set of $SU(2)$ singlet Weyl
fermions and Majorana fermions $N$:
$$  U \sim (1,1,4/3)\quad D \sim (1,1,-2/3)\quad
E \sim(1,1,-2)\quad N \sim (1,1,0).$$ Note the unusual $B-L$
charge assignment of these fields. The BEH bosons, for instance,
have a non-zero $B-L$ charge, and the $N$ are singlet states. They play the role of the heavy Majorana
particles necessary for leptogenesis. 
\begin{figure}[hbt]
\psfig{figure=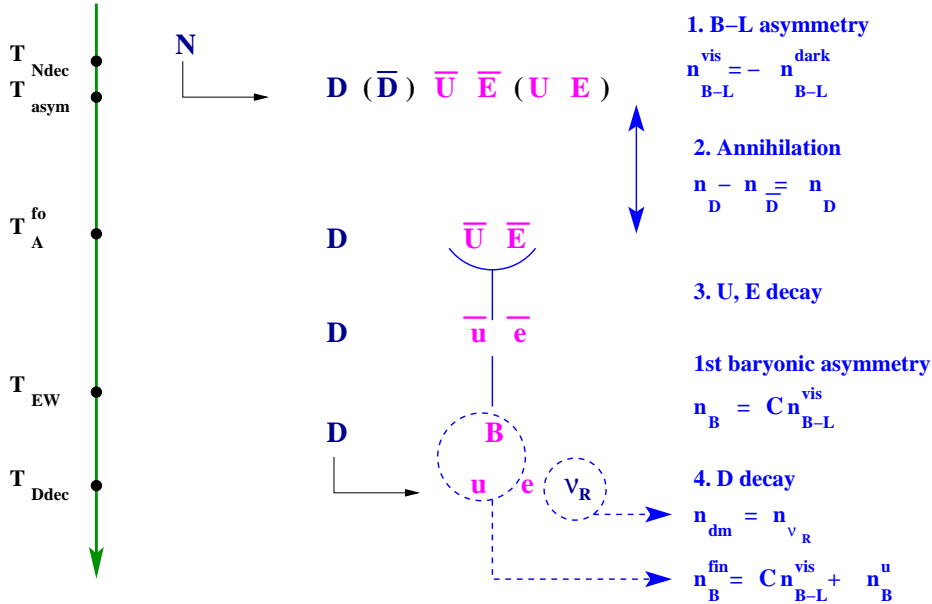,height=8cm}
\caption{Steps of Matter Genesis
\label{fig:radish}}
\end{figure}

The sequence of events is summarized in Figure \ref{fig:radish}.
A $D$ particle excess (dark blue)is produced through the out-of-equilibrium decay of the heavy $N$ states, together 
with an opposite $B-L$ charge in the even-parity, visible, sector (in pink). 
When the universe's temperature reaches $T\sim M_D$, the $D$ and anti-$D$ annihilate each other very efficiently, since they are strongly interacting colored states, leaving
only a tiny $D$ asymmetry. Meanwhile, in the visible sector, $B+L$ violating electroweak processes generate a net baryon asymmetry. Eventually, the relic $D$ particles
decay into the $\nu_R$ plus visible sector particles. If $D$-decay were to take place before electroweak symmetry breaking, the $B-L$ asymmetry 
that was sort of kept apart in the invisible sector would be released free and all $B$ and $L$ excess would be washed out. Hence, an extra constraint is that $D$
decay after electroweak symmetry breaking. The $D$ are thus a bit schizophrenic; they are strongly interacting, heavy $M_D \sim $TeV states 
and yet they must be quite narrow, for they can only decay quite later in the history of the universe,
 $\Gamma_D/M_D \sim 10^{-3}$ (a bit like neutrons). The last constraint we must remember is that the 
 $\nu_R$ can not come in thermal equilibrium, otherwise their thermal abundance would overshadow that produced by the decay of the $D$. The universal see-saw models 
 contains many parameters and, not surprisingly, all our
 constraints can be accommodated by the universal see-saw model with a $D$ particle of mass $\sim $TeV. The details can be found in the original article.\cite{Cosme:2005sb}

\section{Conclusions}

The model presented here is not particularly attractive phenomenologically speaking but it offers nevertheless some interesting features. 
The sort of dark matter we are considering
is by construction very weakly interacting. Hence, unfortunately, it will escape to both direct and indirect detection is any foreseeable future. 
On the other hand there are specific signals that could be seen at the LHC, if the $D$ particles, which
 are basically new quark states, albeit $SU(2)$ singlet ones, are not too heavy.

The dark matter candidate is produced non-thermally through the decay of the $D$ particles. They are thus initially very relativistic. However they have some time to lose their
momentum and their streaming length  is of the order of 1 $M_{pc}$ in the most interesting part of parameters space. The $\nu_R$ are thus a form of
warm rather than cold dark matter \cite{Kitano:2005ge} a feature which could be of interest for structure formation. 

The least satisfying, but inevitable, aspect of the present scenario is that we have to fix by hand the mass of the $\nu_R$ to be 
of the same order as the mass of the light baryons $\sim 1$ GeV. There has been some attempts at circumventing this issue. We would like to mention two such scenarios.
The first one assumes the existence of new, exotic hadronic states \cite{Farrar:2005zd}; the mass scale comes in naturally, the drawback being that we know too much about strong
interactions so that there is little room for speculation... The other attempt invokes the possibility of varying particle masses \cite{Catena:2004pz}. In this
approach baryons
and dark matter masses are function of particle number densities. The main drawback is that it is not easy to reconcile this approach 
with constraints on mass variation.

\section*{Acknowledgments}

The work reported here has been done in collaboration with  Nicolas Cosme and Laura Lopez Honorez.

\end{document}